\documentclass{llncs}

\usepackage{llncsdoc}
\usepackage{graphicx}

\newcommand{\keywords}[1]{\par\addvspace\baselineskip
\noindent\keywordname\enspace\ignorespaces#1}

\begin {document}
\title{Fast Blocked Clause Decomposition with High Quality}
\titlerunning{Fast Blocked Clause Decomposition}

\author{Jingchao Chen}
\institute{School of Informatics, Donghua University \\
2999 North Renmin Road, Songjiang District, Shanghai 201620, P. R.
China \email{chen-jc@dhu.edu.cn}}

\maketitle
\begin{abstract}

Any CNF formula can be decomposed two blocked subsets such that both
can be solved by BCE (Blocked Clause Elimination). To make the
decomposition more useful, one hopes to have the decomposition as
unbalanced as possible. It is often time consuming to achieve this
goal. So far there have been several decomposition and
post-processing algorithms such as \emph{PureDecompose},
\emph{QuickDecompose}, \emph{EagerMover} etc. We found that these
existing algorithms are often either inefficient or low-quality
decomposition. This paper aims at improving the decomposition
quality, while keeping the runtime of algorithms under control. To
achieve this goal, we improve the existing BCE, and present two new
variants of \emph{PureDecompose}, a new heuristic decomposition
called \emph{LessInterfereDecompose}, and a new post-processing
algorithm called \emph{RsetGuidedDecompose}. Combining these new
techniques results in a new algorithm called \emph{MixDecompose}. In
our experiments, there is no application formula where the quality
of \emph{PureDecompose}+\emph{EagerMover} is better than
\emph{MixDecompose}.  In terms of speed, \emph{MixDecompose} is also
very fast. Our average runtime is a
 little longer, but the worst-case runtime is shorter.
 In theory, our two variants of
\emph{PureDecompose} requires linear time in the number of clauses.
By limiting the size of the touch list used by BCE, we can guarantee
always that \emph{MixDecompose} runs in linear time.

\end{abstract}

\section{Introduction}
Recently, one found that blocked clause decomposition (BCD) can not
only efficiently find backbone variables \cite{backbone:97} and
implied binary equivalences through SAT sweeping, but also improve
the performance of the state-of-the-art SAT solvers such as
Lingeling \cite{Lingeling:13} on hard application benchmarks
\cite{sbliter:13,EagerMover:14}. From our experimental result of solver abcdSAT \cite{abcdSAT},
winner of the main track of SAT Race 2015, abcdSAT with BCD was better than abcdSAT without BCD.
 This shows further that BCD is a useful technique.
 Now many researchers have been attracted
to pay attention to this subject.

 A set of clauses is said to be a blocked set if it can be removed
 completely by Blocked Clause Elimination (BCE) \cite{BCE:99,BCE:12}.
 Any CNF formula can be decomposed into two blocked subsets.
To make a blocked clause decomposition more useful, one wants always
to have two blocked subsets as unbalanced as possible. The problem
is that it is not easy to find the most unbalanced subsets. In
theory, one has proven that finding a maximal blocked subset of a
CNF formula with the largest cardinality (\emph{MaxBS} for short) is
NP-hard \cite{sbliter:13}. In other words, it is impossible to find
the best decomposition in polynomial time unless $P = NP$. So far a
few decomposition algorithms were proposed. However, no algorithm
achieves optimization in all terms. \emph{PureDecompose} \cite
{sbliter:13} is the fastest, but its quality is poor. To improve the
quality, Heule et al \cite{sbliter:13} presented
\emph{QuickDecompose}. However, \emph{QuickDecompose} is
time-consuming. Soon after, to improve the speed, Balyo et al
\cite{EagerMover:14} developed a post-processing algorithm called
\emph{EagerMover}. Through an exhaustive series of experiments, we
noted that although the decomposition quality of
\emph{PureDecompose}+\emph{EagerMover} (\emph{PureEager} for short)
and \emph{QuickDecompose} can outperform \emph{PureDecompose}, their
quality is not high yet.

   This paper aims at improving the decomposition quality, while
keeping the runtime of algorithms under control. To achieve this
goal, we present two new variants of \emph{PureDecompose},
 a new decomposition algorithm based
 on clause correlation degree, and a new post-processing algorithm.
 In addition, we improve the existing BCE to speed up the
 decomposition. The algorithm resulting from integrating these new techniques is called
 \emph{MixDecompose}, which can improve significantly the quality of
 decomposition. On application instances,
 the decomposition quality of \emph{MixDecompose} is better than that of \emph{PureEager}.
There is no application formula where the quality of
\emph{PureEager} is better than \emph{MixDecompose}. In terms of
speed, \emph{MixDecompose} is still fast. On average, it took 8.97
seconds on our machine, which is a little slower than
\emph{PureEager} which took 7.41 seconds. However, in the worst case,
\emph{MixDecompose} was faster than \emph{PureEager}. The latter
exceeded 300 seconds in some cases, whereas the former took at most
110 seconds.


\section{Preliminaries}
In this section, we present basic concepts that will be used in
subsequent algorithms for blocked clause decomposition.

\vspace{0.5em}
\noindent \textbf{CNF}. It is short for conjunctive
normal form. A formula in CNF is formulated as a conjunction of
clauses, where each clause is a disjunction of literals, each
literal being either a Boolean variable or its negation. The
negation of a variable $x$ is denoted by $\bar{x}$ or $\neg x$.
 In general, a clause $C$ is written as $C = x_1 \vee
\cdots \vee x_m$, where $x_i (1 \leq i \leq m)$ is a literal. A
formula $F$ is written as $F = C_1 \wedge \cdots \wedge C_n $, where
$C_i (1 \leq i \leq n)$ is a clause. The symbols $var(F)$ and
$lit(F)$ denote the sets of variables and literals occurring in a
formula $F$, respectively.

\vspace{0.5em}

\noindent \textbf{Resolution}. Given two clauses $C_1 = l \vee a_1
\vee \cdots \vee a_m$ and $C_2 = \bar{l} \vee b_1 \vee\cdots \vee
b_n$, the clause $C = a_1 \vee \cdots \vee a_m \vee b_1 \vee \cdots
\vee b_n$ is called the resolvent of $C_1$ and $C_2$ on the literal
$l$,  which is denoted by $C = C_1 {\otimes}_l C_2$.

\vspace{0.5em}

\noindent \textbf{Blocked Clauses}.Given a CNF formula $F$, a clause
$C$, a literal $l \in C$ is said to block $C$ w.r.t. $F$ if (i) $C$
is a tautology w.r.t. $l$,  or (ii) for each clause $C' \in F$ with
$\bar{l} \in C'$, $C' {\otimes}_l C$ is a tautology. A clause is a
tautology if it contains both $x$ and $\bar{x}$ for some variable
$x$. When $l$ blocks $C$ w.r.t. $F$, the literal $l$ and the clause
$C$ are called a blocking literal and a blocked clause,
respectively.

\vspace{0.5em} \noindent \textbf{BCE}. It is short for blocked
clause elimination, which removes blocked clauses from CNF formulas.
By BCE($F$) we mean the CNF formula resulting from repeating the
following operation until fixpoint: If there is a blocked clause $C
\in F$ w.r.t. $F$, let $F := F - \{C\}$. It is said that BCE can
solve a formula $F$ if and only if BCE($F) = \emptyset$. The seminal
work in BCE is due to Kullmmann \cite{BCE:99}.

\section{Blocked Clause Decomposition}

In theory,  any CNF formula can be decomposed into two blocked
subsets. However, not all the decompositions are effective. In
general, The larger one of the blocked sets is, the better the
decomposition quality is, since the larger it is, the more it
resembles the original formula. Therefore, the size difference of
the two sets is considered as a measure of the decomposition
quality. Nevertheless, computing the largest blocked set from a CNF
formula is NP-hard. Hence, we here aim at finding a fast
decomposition with higher quality, rather than the highest quality.

This paper improves \emph{pure decomposition} by defining two
possible variable ordering for variable elimination. The version
based on the ordering from the lowest occurrences to the highest is
called \emph{min pure decomposition}. The version based on the
opposite ordering is called \emph{max pure decomposition}. In
addition, we present a simple and limited BCE and a new
decomposition algorithm, based on this BCE. This new decomposition
algorithm is called \emph{less interfere decomposition}. To improve
further the quality of decomposition, we propose a new
post-processing called \emph{right set guided decomposition}. We do
not know beforehand which one of these algorithms is the best.
However, since these algorithms are lightweight, running all of them
one after the other is fast still. We can obtain a fast and
high-quality algorithm called \emph{MixDecompose} by integrating
sequentially them. In subsequent subsections, we introduce these
algorithms one by one.

\subsection{Pure Decomposition}

This is viewed as the simplest decomposition algorithm. Here we call
it \emph{PureDecompose} for short. Fig. 1 shows its basic idea. Let
the symbols $L$ and $R$ denote the \emph{left}(\emph{large}) subset
and the \emph{right} (\emph{remainder}) subset, respectively. For
each variable $x$, this algorithm adds always the larger of $F_x$
and $F_{\bar x}$ to $L$ and the smaller to $R$, where $F_x$
($F_{\bar x}$) is the set of clauses of $F$ where $x$ occurs
positively (negatively). At the termination of this algorithm, we
have $F = L \cup R $ with $|L| \geq |R|$. In Fig. 1, $\max\{ F_x,
F_{\bar x} \}$ means the set with the larger cardinality between
$F_x$ and $F_{\bar x}$.

   The advantage of this algorithm is that it can be easily implemented
to run in linear time in the size of $F$, using a standard structure
of occurrence lists. Therefore it is very fast. The drawback is that
its decomposition quality is not high on many formulas. For this
reason, next we improve it by defining two possible variable
ordering for variable elimination.

\begin{flushleft}
\begin{sf}
\begin{footnotesize}
\hskip 12mm $PureDecompose (F)$\\
\hskip 16mm $ L := \emptyset $\\
\hskip 16mm {\bf for } each variable $ x \in var(F) $ {\bf do}\\
\hskip 20mm    $L := L \cup \max\{ F_x, F_{\bar x} \}$\\
\hskip 20mm    $F := F - (F_x \cup F_{\bar x})$\\
\hskip 16mm  {\bf return} $L$.

\vspace{1em}

\hskip 8mm \textrm{Fig. 1. Pseudo-code of \emph{PureDecompose}
algorithm}
\end{footnotesize}
\end{sf}
\end{flushleft}

\subsection{Min Pure Decomposition}

   By a few empirical observations, we found that the performance of \emph{PureDecompose} rely significantly
on the order in which variables are eliminated. Here, we present the
first variant of \emph{PureDecompose},  which is called \emph{min
pure decomposition}. Fig. 2 shows its pseudo-code.  Its variable
elimination order is different from \emph{PureDecompose}. One fifth
of variable eliminations are to be eliminated in the same order as
\emph{PureDecompose}. The remaining variables are to be eliminated
in order from the lowest occurrence of literals to the highest. If
there are multiple literals  with the lowest occurrence, the literal
with the minimum total number of clauses containing it is eliminated
first. The total clause size of a literal $x$ can be formulated as
$\sum_{C \in F_x} |C|$.

\begin{flushleft}
\begin{sf}
\begin{footnotesize}
\hskip 12mm $MinPureDecompose (F)$\\
\hskip 16mm $ L := \emptyset $\\
\hskip 16mm $ k := 0 $\\
\hskip 16mm {\bf while } $ F \neq \emptyset $ {\bf do}\\
\hskip 20mm    {\bf if } $k$  $\mathrm{mod}$ $5 = 0 $ {\bf then }
select
$u \in vars(F)$ in the order of variable No. \\
\hskip 20mm    {\bf else} $ m = \min_{x \in lit(F)} |F_x|$\\
\hskip 27mm       $ u :=\arg \min_{|F_x|=m } \sum_{C \in F_x} |C|$\\
\hskip 20mm    $L := L \cup \max\{ F_u, F_{\bar u} \}$\\
\hskip 20mm    $F := F - (F_u \cup F_{\bar u})$\\
\hskip 20mm    $ k := k + 1 $\\
\hskip 16mm  {\bf return} $L$.

\vspace{1em}

\hskip 8mm \textrm{Fig. 2. Pseudo-code of \emph{MinPureDecompose}
algorithm}
\end{footnotesize}
\end{sf}
\end{flushleft}

Compared to \emph{PureDecompose}, this algorithm adds only the
search of variables to be eliminated. This search can be done in
$O(n \log n)$ time, using an order heap, where $n$ is the number of
variables. In the actual implementation, the number $\gamma$ of
variables for computing $\min_{x \in lit(F)} |F_x|$ is limited to
30000 when $n < 70000$, 1500 otherwise. That is, the computation of
$m$ in Fig. 2 is replaced with $ m = \min_{x \leq n \wedge s \leq x
\leq s+ \gamma} \min \{|F_x|,|F_{\bar x}|$\}, where $s$ is the
previous literal $u$ with the lowest occurrence in the given range.
This limit can guarantee that \emph{MinPureDecompose} is still very
fast even if $n$ is very large. In terms of decomposition quality,
this algorithm is superior to the other algorithms on some
application instances such as \emph{ctl\_4291\_567\_5\_unsat\_pre}.

\subsection{Max Pure Decomposition}

Now we consider the second variant of \emph{PureDecompose}. The
order of its variable eliminations is opposite to that of the first
variant. We call this variant \emph{max pure decomposition}, which
is shown in Fig. 3.  It always eliminate first a literal with the
highest occurrence. When multiple literals  have the same highest
occurrence, we select a variable with the lowest difference of its
two literal occurrences. This can be done by computing
$\min_{|F_x|=m} ||F_x| - |F_{\bar x}|| $, where $m$ is defined as
$\max_{x \in lit(F)} |F_x|$. Actually, the first variant can
introduce also this tie-break method.

\begin{flushleft}
\begin{sf}
\begin{footnotesize}
\hskip 12mm $MaxPureDecompose (F)$\\
\hskip 16mm $ L := \emptyset $\\
\hskip 16mm {\bf while } $ F \neq \emptyset $ {\bf do}\\
\hskip 20mm    $ m := \max_{x \in lit(F)} |F_x|$\\
\hskip 20mm    $ u :=\arg \min_{|F_x|=m} ||F_x| - |F_{\bar x}|| $\\
\hskip 20mm    $L := L \cup \max\{ F_u, F_{\bar u} \}$\\
\hskip 20mm    $F := F - (F_u \cup F_{\bar u})$\\
\hskip 16mm  {\bf return} $L$.

\vspace{1em}
\hskip 8mm \textrm{Fig. 3. Pseudo-code of
\emph{MaxPureDecompose} algorithm}
\end{footnotesize}
\end{sf}
\end{flushleft}

   Unlike the first variant, \emph{MaxPureDecompose} needn't compute
the total clause size $\sum_{C \in F_x} |C|$ for each literal $x$.
So it should run faster than the first variant.
   In order to ensure that is still very fast even if the number of variables is very large,
the number $\gamma$ of variables for finding a literal with the
highest occurrence is limited to 5000 when $n < 800000$, 500
otherwise. In other words, in the actual implementation, the
computation of $m$ in Fig. 3 is replaced with $ m = \max_{x \leq n
\wedge s \leq x \leq s+ \gamma} \max \{|F_x|,|F_{\bar x}|$\}, where
$s$ is the previous literal $u$ with the highest occurrence. The
decomposition quality of this algorithm is superior to that of the
other algorithms on some
   application instances such as \emph{complete-500}.

\subsection{A Simple and Limited BCE}

BCE is applied not only to BCD, but also to CNF preprocessing. Using
a good BCE is very important.  To improve the efficiency of
\emph{LessInterfereDecompose} that will be given in the next
subsection, in Fig. 4 we present a simple and efficient BCE, which
is different from that presented in \cite{BCE:12}. BCE in
\cite{BCE:12} is based on a literal-based priority queue, while our
BCE is based on a clause-based linear linked list. Another important
difference from the usual BCE is that we do not try to test whether
each literal $l$ in $C$ is a blocking literal when $|F|\geq300000$.
We test only literal $l$ with $|F_{\bar l}|<2$. That is, we replace
the statement `` {\sf{\bf for }  $ l \in C $ {\bf do} }" in the
usual BCE with the statement `` {\sf {\bf for } $ l \in C $ with
($|F_{\bar l}|<2$ or $|F|<300000$ or \emph{isFirst}) {\bf do}}'',
where \emph{isFirst} is a Boolean variable for testing whether BCE
is invoked for the first time. If it is the first time to call to
BCE, we run the usual BCE. The condition ``$|F_{\bar l}|<2$'' will
not prevent forever from calling to BCE, since we select always
literal $\bar l$ with the minimum occurrences, and move at least one
clause from $F_{\bar l}$ to $R$ every time, i.e., $|F_{\bar l}|$
decreases constantly.

 In the decomposition algorithm given in the next subsection, using this simple BCE is
much faster than the usual BCE. Surprisingly, the decomposition
quality keep unchanged in most cases. Even if it is changed, its
change is still very small. In addition, our \emph{ touch} function
is different from that in \cite{BCE:12}. Here is our definition
about it.
\[{\it touch(C,F)}= \left \{ \begin {array}
         {l@{\quad \quad}l}
         \bigcup\limits_{x \in C}F_{\bar x} & |F| < 800000 \hskip 1mm \mathrm{or} \hskip 1mm isFirst \\
         \bigcup\limits_{x \in C \wedge |F_x|<2}F_{\bar x} & \mathrm
         {otherwise}
         \end {array} \right . \]\\
When $|F| \geq 800000$, and it is not the first call to $BCE$, we
consider only the clauses touched by the negation of literals with
the number of occurrences $< 2$. This can speed up the decomposition
of large instances. For example, using the above \emph{touch},  the
runtime required by \emph{LessInterfereDecompose} to decompose
\emph{q\_query\_3\_L90} can be reduced from 600 seconds to less than
9 seconds. The decomposition quality keep unchanged still.

\begin{flushleft}
\begin{sf}
\begin{footnotesize}

\hskip 12mm $BCE($touched clauses $T$, formula $F$, blocked set $L)$\\
\hskip 16mm {\bf for } each clause $ C \in T \wedge F$ {\bf do}\\
\hskip 20mm    {\bf for }  $ l \in C $ with ($|F_{\bar l}|<2$ or $|F|<300000$ or \emph{isFirst}) {\bf do}\\
\hskip 24mm    {\bf if } all resolvents of $C$ on $l$ are tautologies, i.e., $C$ is blocked {\bf then}\\
\hskip 28mm $L := L \cup \{C\}$\\
\hskip 28mm $F := F - \{C\}$\\
\hskip 28mm $T := T \cup touch(C, F)$\\
\hskip 28mm {\bf continue} with next $C$ in outer loop\\
\hskip 16mm {\bf return} $L$

\vspace{1em}
\hskip 8mm \textrm{Fig. 4. Pseudo-code of \emph{BCE}
algorithm}
\end{footnotesize}
\end{sf}
\end{flushleft}

\subsection{Less Interfere Decomposition}

The two variants of \emph{PureDecompose} both are based on the
variable elimination order. Nevertheless, in fact, it is difficult
to find out the optimal algorithm by optimizing only the variable
elimination order. Therefore, it is necessary to find a different
decomposition technique. Below we present a new algorithm called
\emph{less interfere decomposition}, which is based on the order of
clause elimination. Fig. 5 shows its pseudo-code. The basic outline
of this algorithm may be sketched as follows: move blocked clauses
in $F$ to $L$ by BCE, compute the candidate set $S$, move each
clause $C \in S \cap F$ to $R$. These steps are repeated until $F$
is empty. The computation of the candidate set $S$ is based on the
notion of interfering degree. The interfering degree of a clause $C$
can be defined as $\sum\limits_{C' \in F \wedge l \in C \wedge
\bar{l} \in C'} Ntaut(C' {\otimes}_l C) $, where $Ntaut(X)$ is zero
if $X$ is a tautology clause, and one otherwise. The probability
that $C' {\otimes}_l C$ is not a tautology clause is very high. To
save the computing cost, we may approximate the interfering degree
as $\sum\limits_{C' \in F} |\{l \mid l \in C \wedge \bar{l} \in
C'\}|$. \emph{LessInterfereDecompose} in Fig. 5 uses this
approximation version to compute the interfering degree, and call
this measure $score$, i.e., $score[C]= \sum\limits_{C' \in F} |\{l
\mid l \in C \wedge \bar{l} \in C'\}|$. To get clauses with the
maximum score, all the clauses in $F$ are traversed. If only one
clause with the maximum score is moved to $R$ from $F$ each time $F$
is traversed, it is time consuming. So we decide to move $p$ clauses
to $R$ one time, where $p=\frac{|F|}{\theta}$, where $\theta$ is a
constant. Table\,1 shows the performance behavior for different
$\theta$'s on ACG-20-5p1 with $|F|=1416850$. For this instance,
selecting 400 as the value of $\theta$ is a better choice. However,
considering the other instances, actually for application instances,
$\theta$ is set to 200 when $|F| \geq 8\times10^5$, 2300 Otherwise.
For large instances, because we put great stock in speed, $\theta$
is selected as a smaller value. For small instances, because we put
great stock in quality, $\theta$ is selected as a larger value. This
is actually a compromise between speed and quality. For random
instances, $\theta$ is set to 400 in any case. When
$\frac{|F|}{\theta} < 18 $, $p$ is set to 18. As shown in Fig. 5,
the clauses with the first $p$ highest scores are stored in $S$ as
the candidate clauses to be moved to $R$. In order to save time
further, we compute the interfering degree produced by only literals
with the lowest occurrence, not all literals.

\begin{flushleft}
\begin{sf}
\begin{footnotesize}

\hskip 12mm $LessInterfereDecompose(F)$\\
\hskip 16mm $ L := R := S := \emptyset $\\
\hskip 16mm $BCE(F,F,L)$\\
\hskip 16mm {\bf while } $ F \neq \emptyset $ {\bf do}\\
\hskip 20mm    {\bf if } $ S \cap F = \emptyset $ {\bf then} \\
\hskip 25mm     $ m = \min_{x \in lit(F)} |F_x|$\\
\hskip 25mm        {\bf for } each clause $ C \in F ${\bf do}\\
\hskip 28mm           {\bf for } each clause $e \in F_l$ with $ l \in C $ and $|F_l|=m$ {\bf do}\\
\hskip 32mm              $ score[e] := score[e]+1$\\
\hskip 25mm         $S :=$ \{$x | score[x] \geq \alpha$, where the $p$-th highest score is $\alpha$\}\\
\hskip 20mm    select a clause $C \in S \cap F $\\
\hskip 20mm    $F := F - \{C\}$\\
\hskip 20mm    $BCE(touch(C, F),F, L)$\\
\hskip 16mm {\bf return} $L$

\vspace{1em} \hskip 8mm \textrm{Fig. 5. Pseudo-code of
\emph{LessInterfereDecompose} algorithm}
\end{footnotesize}
\end{sf}
\end{flushleft}

\begin{table}
\caption{ Performance of \emph{LessInterfereDecompose} +
post-processing given in Fig.7 for different $\theta$'s on
ACG-20-5p1. Time is in seconds}

\begin{center}

\setlength\tabcolsep{4pt}
\begin{tabular}{|r|r|c|}
\hline  \hline \ $\theta$ & Time & $\frac{|L|}{|F|}$ \\
 \hline
50  & 6.95 &  79.66\% \\
100 & 6.03 &  78.87\% \\
200 & 5.44 &   79.11\% \\
400 & 5.57 &   79.32\% \\
800 & 6.52 &  79.31\% \\
1600 &   8.62  &  79.77\% \\
2400 &   10.64  & 79.78\% \\

 \hline
\end{tabular}
\end{center}
\end{table}

\nopagebreak[3]

   The runtime of \emph{LessInterfereDecompose} consists of three
   parts: \emph{BCE}, computing scores and determining $S$'s. The runtime of
computing scores is $O(\frac{|F|^2}{p})=O(\theta|F|)$. If scores are
given, determining a $S$ can be done in a linear time in $|F|$,
since there exist linear time algorithms for finding the $p$-th
highest score \cite{find:97,find:00}. The total runtime of computing
scores plus determining $S$'s does not exceed $O(\theta|F|)$. If the
number of clauses touched by each clause does not exceed a constant
$\delta$, where $\delta$  is certainly smaller than the maximal
number of literal occurrences times the maximal size of clauses,
i.e., $\max_{x \in lit(F)} |F_x| \times \max_{C \in
 F}|C|$. The total time required by all \emph{BCE}'s is at most $O(\delta
|F|)$. Thus, the total runtime of \emph{LessInterfereDecompose} is
at most $O((\delta+\theta)|F|)$. In practice, $\delta$ is generally
very small. Should $\delta$ is very large, we can remove a part of
touched clauses to reduce the time required by \emph{BCE} to test
whether a clause in the touch list is blocked, or limit the size of
the touch list a small constant, say 2000. Using such a policy can
guarantee that the time complexity of \emph{LessInterfereDecompose}
is linear in $|F|$. Compared with \emph{EagerMover} in
\cite{EagerMover:14}, the runtime of \emph{BCE} in
\emph{LessInterfereDecompose} is smaller than that in
\emph{EagerMover}.  \emph{EagerMover} calls at least four times
\emph{BCE} on a subset with the size of $0.75|F|$. All calls to
\emph{BCE} on each clause $C$ in $F$ can be viewed as a call to
\emph{BCE} on the whole $F$. Thus, the total runtime of \emph{BCE}
in \emph{LessInterfereDecompose} corresponds to double the runtime
of \emph{BCE} on a $F$. As long as the runtime of computing scores
and determining $S$'s is smaller than the runtime of \emph{BCE} on a
$F$, \emph{LessInterfereDecompose} should be faster than
\emph{EagerMover}. In fact, that is true. On some instances, the
former are indeed faster than the latter.

\subsection{Right Set Guided Post-processing}

 In general, the above
algorithms do not achieve maximal blocked set decomposition.
However, they can be improved further by post-processing. The
post-processing often used is \emph{MoveBlockedClause} algorithm
shown in Fig. 7, which is to move blocked (with respect to the
current $L$) clauses from $R$ to $L$. We noted that even if this
post-processing algorithm is applied, the decomposition quality can
be improved still. For this reason, we present a new post-processing
algorithm, called \emph{Right set guided decomposition}, which is
shown in Fig. 6.  It is a simplified version of
\emph{LessInterfereDecompose}. Replacing $S$ with $R$ results in
this algorithm. This algorithm requires that the right blocked  set
$R$ must be given in advance. Hence, it is used generally as
post-processing.  It is faster than \emph{LessInterfereDecompose},
since it need not compute $R$. Its time complexity depends mainly
on that of \emph{BCE}. For some benchmarks, this algorithm can
improve significantly their decomposition quality. For example, to
decompose \emph{SAT\_dat.k75-24\_1\_rule\_3} using
\emph{MinPureDecompose}, the fractions of the large subset (i.e.,
$\frac{|L|}{|F|}$) with \emph{RsetGuidedDecompose} and without it
are $83.9\%$ and $69.9\%$, respectively. If replacing
\emph{MinPureDecompose} with \emph{LessInterfereDecompose}, their
fractions are 87.8\% and 87.3\%, respectively.
\emph{RsetGuidedDecompose} raises still the quality by 0.5\%.
However, the speed difference among the three algorithms is big. On
this instance, \emph{LessInterfereDecompose},
\emph{RsetGuidedDecompose} and \emph{MinPureDecompose} spent 25, 4
and 1 seconds, respectively. The slowest
\emph{LessInterfereDecompose} is not suitable for huge instances
with ten millions of clauses.

\begin{flushleft}
\begin{sf}
\begin{footnotesize}

\hskip 12mm $RsetGuidedDecompose($formula $F$, right set $R$)\\
\hskip 16mm $ L := \emptyset $\\
\hskip 16mm $BCE(F,F,L)$\\
\hskip 16mm {\bf while } $ F \neq \emptyset $ {\bf do}\\
\hskip 20mm    select a clause $C \in (R \cap F) $\\
\hskip 20mm    $F := F - \{C\}$\\
\hskip 20mm    $BCE(touch(C, F),F, L)$\\
\hskip 16mm {\bf return} $L$

\vspace{1em}

\hskip 8mm \textrm{Fig. 6. Pseudo-code of \emph{RsetGuidedDecompose}
algorithm}
\end{footnotesize}
\end{sf}
\end{flushleft}

\subsection{Mix Decomposition}

\begin{flushleft}
\begin{sf}
\begin{footnotesize}
\hskip 22mm $MoveBlockedClause($left blocked set $L$, right set $R)$\\
\hskip 26mm {\bf for } each clause $ C \in R$ {\bf do}\\
\hskip 30mm    {\bf if } $\mathrm{BCE}(L \cup \{C\})=\emptyset$ {\bf then} $L := L \cup \{C\}$\\
\hskip 26mm {\bf return} $L$

\vspace{1em}

\hskip 12mm $MixDecompose($formula $F$)\\
\hskip 16mm $L_1:=PureDecompose(F)$\\
\hskip 16mm $L_2:=MinPureDecompose(F)$\\
\hskip 16mm $L_3:=MaxPureDecompose(F)$\\
\hskip 16mm $L:=\max\{L_1,L_2,L_3\}$\\
\hskip 16mm  {\bf if } $|F|<5\times10^6$ and $|var(F)|<10^6$ {\bf then} \\
\hskip 20mm    $L_4:=LessInterfereDecompose(F)$\\
\hskip 20mm    $L:=\max\{L,L_4\}$\\
\hskip 16mm $L:=RsetGuidedDecompose(F, F-L)$\\
\hskip 16mm $L:=MoveBlockedClause(L,F-L)$\\
\hskip 16mm {\bf return} $L$

\vspace{1em}

\hskip 8mm \textrm{Fig. 7. Pseudo-code of \emph{MoveBlockedClause},
\emph{MixDecompose} algorithm}
\end{footnotesize}
\end{sf}
\end{flushleft}

In general, in advance we do not know which algorithm is the best.
Because all the algorithms given in the previous subsections are
very fast, and running them one after another does not lose much
time, we can construct an algorithm with high speed and high
performance by combining them. The detailed implementation is shown
in Fig. 7. We call this algorithm \emph{MixDecompose}. Its basic
idea is to take the maximum from three left sets outputted by three
algorithms as the initial $L$ first. If the formula to be decomposed
is not large, say the number of clauses and variables is less than
$5\times10^6$ and $10^6$, respectively, we invoke
\emph{LessInterfereDecompose} to get a larger $L$. Finally, we
enlarge the size of $L$ by calling two post-processing algorithms:
\emph{RsetGuidedDecompose} and \emph{MoveBlockedClause}. Like the
usual post-processing, the task of \emph{MoveBlockedClause} is to
move blocked clauses from $R$ to $L$. Notice, if $F$ is large, say
$|F| > 10^7$, the last post-processing can be canceled to save the
running time.

According to whether both subsets can be solved by BCE, a blocked
clause decomposition can be classified as \emph{symmetric} or
\emph{asymmetric}. If yes, it is symmetric. If only one of the
subsets can be solved by BCE, it is asymmetric. Clearly, our two
variants of \emph{PureDecompose} are symmetric. If blockable clauses
(whose definition is given below) are allowed to move to $L$ like
\emph{EagerMover} \cite{EagerMover:14}, that is, replacing
\emph{MoveBlockedClause} with the procedure
\emph{MoveBlockableClause} of \emph{EagerMover},
 \emph{MixDecompose} is asymmetric,
since it cannot guarantee that the two subsets both can be solved by
BCE. However, even if adopting the replacement, by our observation,
on almost all application instances, it is still symmetric.

\section{Empirical Evaluation}

  We evaluated the performance of each decomposition algorithm on the
297 instances from the application track of the SAT competition
2014, except for three huge ones: zfcp-2.8-u2-nh,
esawn\_uw3.debugged and post-cbmc-zfcp-2.8-u2. The reason why the
three huge instances were removed is that there is not enough memory
to solve them. All the algorithms were run under the following
experimental platform: Intel Core 2 Quad Q6600 CPU with speed of
2.40GHz and 2GB memory. Each tested algorithm is written in C. The
source code of \emph{MixDecompose} is available at
http://github.com/jingchaochen/MixBcd.

  This paper presented four decomposition algorithms. To
understand more clearly the characteristic of each of them, we
compared them experimentally. Empirical results reveal that except
for 41 application instances listed in Table\,2, on all the other
ones, the quality of \emph{LessInterfereDecompose} is superior to
that of the other three algorithms: \emph{MinPureDecompose},
\emph{PureDecompose}, \emph{MaxPureDecompose}. That is to say, there
are 256 application instances where \emph{LessInterfereDecompose} is
superior to the other three algorithms in terms of quality. However,
due to limited space, Table\,3 lists only a part of instances. In
Table\,2--4, $|F|$ denotes the number of the clauses in formula $F$,
where $F$ is simplified by removing satisfied clauses, but contains
unit clauses. To obtain such $F$, before calling each decomposition
algorithm, we use the same unit decomposition policy given in
\cite{EagerMover:14} as \emph{EagerMover} to preprocess the input
formula. Column $\frac{|L|}{|F|}$ indicates the fraction of the
large set. Column Time shows the runtime in seconds.

  Only from Table\,2, \emph{MinPureDecompose} seems to be less important than the
  others, since there are only two instances where
  it is better than the others. However, in fact, on some large
  instances, it is very important. For example, for
  \emph{9vliw\_m\_9stage\_iq3\_C1\_b1}, it is very important, because on this instance, \emph{LessInterfereDecompose}
is much slower than \emph{MinPureDecompose}, but their quality
difference is small, as shown in the last row of Table\,3. In
\emph{MixDecompose} execution, to save the runtime, we skip
\emph{LessInterfereDecompose} and adopt the best result of the other
algorithms (which may well be \emph{MinPureDecompose}) when $|F|
\geq 5\times10^6$.

\begin{table}
\caption{All application instances where
\emph{LessInterfereDecompose} (\emph{LessInterfere} for short) is
inferior to the other three algorithms:\emph{MinPureDecompose}
(\emph{MinPure} for short), \emph{PureDecompose},
\emph{MaxPureDecompose} (\emph{MaxPure} for short). Time is in
seconds.}
\begin{center}

\setlength\tabcolsep{4pt}
\begin{tabular}{|l|r|c|c|c|c|c|c|c|c|}
\hline  \hline
\  & & \multicolumn{2}{c|} {\emph{MinPure}} &  \multicolumn{2}{c|}{\emph{PureDecompose}} & \multicolumn{2}{c|} {\emph{MaxPure}} &  \multicolumn{2}{c|}{\emph{LessInterfere}} \\
\cline{3-10} \multicolumn{1}{|c|}{\raisebox{1.5ex}[0pt]{Instances}}
  & \raisebox{1.0ex}[0pt]{\large $\frac{|F|}{10^4}$} & $\frac{|L|}{|F|}$ & Time & $\frac{|L|}{|F|}$ & Time & $\frac{|L|}{|F|}$ & Time & $\frac{|L|}{|F|}$ & Time\\
 \hline
ctl\_3791\_556\_unsat    & 8  & \textbf{93.1\%} & 0.14 & 88.9\% & 0.01 & 75.3\% & 0.03 & 88.1\% & 3.17\\
ctl\_4291\_567\_5\_unsat & 13 & \textbf{87.6\%} & 0.66 & 83.7\% & 0.01 & 68.6\% & 0.04 & 80.3\% & 9.94\\

atco\_enc1\_opt2\_20\_12 & 653 & 78.5\% & 1.48 & \textbf{82.9\%}  & 0.51  & 77.3\% & 1.47 & 80.3\% & 57.7 \\
atco\_enc2\_opt1\_20\_11 & 971 & 84.2\% & 2.34 & \textbf{87.0\%}  & 0.86  & 82.7\% & 3.75 & 86.0\% & 64.2 \\
atco\_enc2\_opt2\_20\_11 & 651 & 78.7\% & 1.40 & \textbf{83.0\%}  & 0.50  & 77.3\% & 1.46 & 80.4\% & 58.1 \\
atco\_enc3\_opt1\_03\_53 & 427 & 51.0\% & 6.61 & \textbf{75.8\%}  & 0.74  & 75.4\% & 12.9 & 60.5\% & 35.2 \\
atco\_enc3\_opt1\_04\_50 & 561 & 50.5\% & 8.90 & \textbf{75.1\%}  & 1.01  & 75.0\% & 21.9 & 56.9\% & 60.2 \\
atco\_enc3\_opt1\_13\_48 & 608 & 50.5\% & 9.92 & \textbf{75.1\%}  & 1.12  & 75.0\% & 22.8 & 56.9\% & 70.4 \\
atco\_enc3\_opt2\_05\_21 & 538 & 50.8\% & 8.63 & \textbf{75.8\%}  & 1.02  & 75.7\% & 21.3 & 57.8\% & 55.2 \\

grieu-vmpc-31            &  15  & 79.7\% & 0.03 & \textbf{79.8\%} & 0.01 & 79.7\% & 0.01 & 79.7\% & 4.60  \\
openstack-p30\_3.085     &  141 & 76.7\% & 0.52 & \textbf{92.8\%} & 0.10 & 85.1\% & 0.13 & 89.6\% & 2.04  \\
openstack-s-p30\_3.085   &  141 & 76.7\% & 0.51 & \textbf{92.8\%} & 0.11 & 85.1\% & 0.14 & 89.6\% & 2.03  \\
reg\_s\_2\_unknown       &  170 & 77.0\% & 1.17 & \textbf{79.4\%} & 0.16 & 69.0\% & 1.22 & 72.3\% & 101   \\

vmpc\_29                 & 12   & 79.6\% & 0.01 & \textbf{79.7\%} & 0.01 & 79.6\% & 0.01 & 79.6\% & 4.17 \\
vmpc\_32                 & 16   & 79.6\% & 0.02 & \textbf{79.7\%} & 0.01 & 79.6\% & 0.01 & 79.6\% & 5.59 \\
vmpc\_33                 & 18   & 79.6\% & 0.04 & \textbf{79.7\%} & 0.01 & 79.6\% & 0.01 & 79.6\% & 6.81 \\

atco\_enc3\_opt2\_10\_12 & 422  & 50.3\% & 6.61 & 75.0\% & 0.77 & \textbf{75.1\%} & 14.2 & 60.0\% & 35.7 \\
atco\_enc3\_opt2\_10\_14 & 423  & 50.3\% & 6.64 & 75.0\% & 0.78 & \textbf{75.1\%} & 14.2 & 60.0\% & 36.1 \\
atco\_enc3\_opt2\_18\_44 & 457  & 50.3\% & 7.22 & 75.0\% & 0.79 & \textbf{75.1\%} & 14.5 & 60.0\% & 41.8 \\

complete-300-0.1-18      & 3    & 90.2\% & 0.01 & 90.0\% & 0.01 & \textbf{93.8\%} & 0.01 & 90.8\% & 0.79 \\
complete-300-0.1-4       & 3    & 90.8\% & 0.01 & 89.3\% & 0.01 & \textbf{92.9\%} & 0.01 & 90.4\% & 0.72 \\
complete-300-0.1-7       & 3    & 90.2\% & 0.01 & 89.8\% & 0.01 & \textbf{93.5\%} & 0.01 & 90.8\% & 0.81 \\
complete-300-0.1-8       & 3    & 90.8\% & 0.01 & 90.2\% & 0.01 & \textbf{93.0\%} & 0.01 & 90.6\% & 0.71 \\
complete-400-0.1-12      & 5    & 92.8\% & 0.01 & 91.6\% & 0.01 & \textbf{94.6\%} & 0.01 & 92.0\% & 2.70 \\
complete-400-0.1-16      & 5    & 91.5\% & 0.01 & 91.7\% & 0.02 & \textbf{94.5\%} & 0.01 & 92.8\% & 2.71 \\
complete-400-0.1-3       & 5    & 91.9\% & 0.01 & 92.0\% & 0.02 & \textbf{94.7\%} & 0.01 & 91.8\% & 2.65 \\
complete-400-0.1-7       & 5    & 91.7\% & 0.02 & 91.3\% & 0.01 & \textbf{94.6\%} & 0.01 & 92.1\% & 2.61 \\
complete-500-0.1-1       & 7    & 92.9\% & 0.03 & 93.0\% & 0.03 & \textbf{95.5\%} & 0.01 & 93.3\% & 1.97 \\
complete-500-0.1-15      & 7    & 92.7\% & 0.03 & 93.3\% & 0.03 & \textbf{95.5\%} & 0.01 & 92.6\% & 2.11 \\
complete-500-0.1-17      & 7    & 93.3\% & 0.04 & 93.1\% & 0.04 & \textbf{95.6\%} & 0.01 & 92.8\% & 2.08 \\
complete-500-0.1-7       & 7    & 93.2\% & 0.02 & 93.1\% & 0.03 & \textbf{95.6\%} & 0.01 & 93.3\% & 2.01 \\
complete-500-0.1-8       & 7    & 93.5\% & 0.03 & 92.8\% & 0.02 & \textbf{95.7\%} & 0.01 & 93.3\% & 1.98 \\

pb\_300\_10\_lb\_07      & 55   & 58.0\% & 0.70 & 64.7\% & 0.02 & \textbf{75.5\%} & 0.10 & 71.0\% & 15.1 \\
pb\_300\_10\_lb\_08      & 56   & 58.0\% & 0.71 & 64.7\% & 0.04 & \textbf{75.5\%} & 0.09 & 70.6\% & 15.3 \\

stable-300-0.1-20        & 2    & 88.9\% & 0.01 & 89.6\% & 0.01 & \textbf{93.3\%} & 0.01 & 87.1\% & 0.52 \\
stable-400-0.1-11        & 3    & 90.5\% & 0.01 & 91.8\% & 0.01 & \textbf{94.4\%} & 0.01 & 89.8\% & 1.51 \\
stable-400-0.1-12        & 3    & 91.2\% & 0.02 & 90.4\% & 0.01 & \textbf{94.1\%} & 0.01 & 89.2\% & 1.60 \\
stable-400-0.1-2         & 3    & 90.6\% & 0.01 & 90.5\% & 0.01 & \textbf{94.3\%} & 0.01 & 88.9\% & 1.59 \\
stable-400-0.1-4         & 3    & 90.3\% & 0.01 & 91.2\% & 0.01 & \textbf{94.3\%} & 0.01 & 89.1\% & 1.63 \\
stable-400-0.1-5         & 3    & 89.5\% & 0.01 & 91.6\% & 0.01 & \textbf{94.1\%} & 0.01 & 88.7\% & 1.61 \\
stable-400-0.1-7         & 3    & 89.7\% & 0.02 & 91.3\% & 0.01 & \textbf{94.3\%} & 0.01 & 89.0\% & 1.56 \\

 \hline
\end{tabular}
\end{center}
\end{table}

\begin{table}
\caption{Some application instances where the quality of
\emph{LessInterfereDecompose}(\emph{LessInterfere}) is superior to
that of the other three algorithms: \emph{MinPureDecompose}
(\emph{MinPure}), \emph{PureDecompose}, \emph{MaxPureDecompose}
(\emph{MaxPure}). Time is in seconds.}
\begin{center}

\setlength\tabcolsep{4pt}
\begin{tabular}{|l|r|c|c|c|c|c|c|c|c|}
\hline  \hline
\  & & \multicolumn{2}{c|} {\emph{MinPure}} &  \multicolumn{2}{c|}{\emph{PureDecompose}} & \multicolumn{2}{c|} {\emph{MaxPure}} &  \multicolumn{2}{c|}{\emph{LessInterfere}} \\
\cline{3-10} \multicolumn{1}{|c|}{\raisebox{1.5ex}[0pt]{Instances}}
  & \raisebox{1.0ex}[0pt] {\large $\frac{|F|}{10^4}$} & $\frac{|L|}{|F|}$ & Time & $\frac{|L|}{|F|}$ & Time & $\frac{|L|}{|F|}$ & Time & $\frac{|L|}{|F|}$ & Time\\
 \hline

001-80-12                &  31  &  58.3\% & 0.97 &  57.7\% & 0.03 & 57.5\% & 0.06 & 99.2\% &  6.05 \\
010-80-12                &  31  &  58.3\% & 0.94 &  57.7\% & 0.04 & 57.5\% & 0.05 & 99.2\% &  6.09 \\
6s10                     &  10  &  66.5\% & 3.18 &  63.0\% & 0.01 & 63.0\% & 0.12 & 99.9\% &  0.07 \\
6s123                    &  241 &  63.4\% & 4.99 &  62.5\% & 0.25 & 60.6\% & 0.17 & 97.7\% &  0.67 \\
7pipe\_k                 &  75  &  96.7\% & 1.91 &  96.0\% & 0.12 & 96.1\% & 0.04 & 99.9\% &  9.68 \\
8pipe\_k                 &  133 &  97.2\% & 1.58 &  96.7\% & 0.27 & 96.7\% & 0.09 & 97.9\% &  2.74 \\
9dlx\_vliw\_at\_b\_iq3   &  97  &  93.4\% & 4.93 &  92.7\% & 0.12 & 92.2\% & 0.11 & 96.7\% &  1.21 \\
9dlx\_vliw\_at\_b\_iq9   & 968  &  95.1\% & 3.52 &  94.4\% & 1.61 & 94.3\% & 3.26 & 96.5\% &  50.5 \\
ACG-15-10p0              &  92  &  71.5\% & 1.06 &  71.8\% & 0.12 & 68.7\% & 2.16 & 76.3\% &  1.93 \\
aes\_24\_4\_keyfind\_4   &  1   &  54.4\% & 0.01 &  54.3\% & 0.01 & 55.1\% & 0.01 & 66.9\% &  0.21 \\
aes\_64\_1\_keyfind\_1   & 0.3  &  52.8\% & 0.01 &  56.1\% & 0.01 & 50.7\% & 0.01 & 89.2\% &  0.01 \\
AProVE07-01              & 2    &  61.3\% & 0.30 &  61.1\% & 0.01 & 61.1\% & 0.01 & 96.5\% &  0.01 \\
AProVE09-06              & 26   &  62.2\% & 0.57 &  62.2\% & 0.02 & 62.2\% & 0.23 & 99.9\% &  0.16 \\
atco\_enc1\_opt1\_04\_32 & 55   &  69.4\% & 3.08 &  70.7\% & 0.04 & 69.5\% & 0.21 & 76.3\% &  44.3 \\
beempgsol2b1             & 8    &  65.8\% & 2.10 &  62.6\% & 0.11 & 63.1\% & 0.09 & 99.9\% &  0.30 \\
bjrb07amba10andenv       & 59   &  66.7\% & 1.31 &  66.4\% & 0.08 & 66.3\% & 0.94 & 99.9\% &  4.79 \\
blocks-blocks-37-1.130   & 728  &  88.6\% & 1.88 &  90.6\% & 0.60 & 87.4\% & 5.25 & 94.4\% &  9.16 \\
bob12m04                 & 168  &  66.7\% & 3.64 &  60.9\% & 0.17 & 61.5\% & 3.12 & 99.9\% &  2.74 \\
c10bi\_i                 &  40  &  66.7\% & 0.99 &  62.1\% & 0.03 & 62.0\% & 0.41 & 99.9\% &  0.92 \\
countbitssrl032          &  6   &  66.7\% & 1.81 &  58.5\% & 0.01 & 62.7\% & 0.06 & 99.9\% &  0.20 \\
dated-10-11-u            &  48  &  69.4\% & 0.42 &  68.2\% & 0.05 & 62.1\% & 1.05 & 81.4\% &  2.60 \\
dimacs                   &  1   &  58.7\% & 0.01 &  58.9\% & 0.01 & 59.0\% & 0.01 & 99.9\% &  0.01 \\
E02F22                   &  130 &  99.0\% & 1.35 &  98.8\% & 0.29 & 98.2\% & 0.12 & 99.9\% &  28.6 \\
grid-strips-grid-y-3.065 & 350  &  87.1\% & 0.68 &  85.8\% & 0.29 & 90.5\% & 0.28 & 97.1\% &  1.77 \\
gss-25-s100              &  10  &  65.0\% & 3.01 &  64.8\% & 0.01 & 64.6\% & 0.17 & 99.8\% &  0.04 \\
hitag2-8-60-0--47        &  3   &  53.8\% & 0.30 &  52.1\% & 0.01 & 53.3\% & 0.01 & 98.4\% &  0.53 \\
hwmcc10-k45-pdts3p02     & 49   &  66.7\% & 1.09 &  65.2\% & 0.04 & 64.8\% & 0.71 & 99.9\% &  0.25 \\
itox\_vc1130             & 44   &  54.3\% & 0.71 &  54.9\% & 0.03 & 54.7\% & 0.78 & 97.5\% &  1.34 \\
k2fix\_gr\_rcs\_w9.shuffled & 31   &  99.6\% & 0.26 &  99.6\% & 0.06 & 99.6\% & 0.01 & 99.7\% &  0.23 \\
korf-17                  & 9    &  92.9\% & 0.22 &  93.0\% & 0.01 & 90.5\% & 0.04 & 99.5\% &  0.16 \\
manol-pipe-c10nidw       & 129  &  66.7\% & 3.11 &  61.9\% & 0.14 & 61.7\% & 1.43 & 99.9\% &  0.41 \\
maxxor032                & 4    &  66.6\% & 1.06 &  60.9\% & 0.01 & 60.8\% & 0.20 & 99.9\% &  0.04 \\
MD5-32-1                 & 7    &  56.6\% & 0.34 &  51.5\% & 0.01 & 53.3\% & 0.01 & 99.0\% &  0.26 \\
minandmaxor128           & 75   &  66.7\% & 1.61 &  62.1\% & 0.07 & 62.1\% & 0.48 & 99.9\% &  3.14 \\
partial-10-17-s          & 118  &  70.5\% & 1.05 &  69.0\% & 0.14 & 63.2\% & 3.62 & 78.3\% &  2.48 \\
post-c32s-ss-8           & 14   &  62.7\% & 4.06 &  58.3\% & 0.01 & 57.9\% & 0.15 & 96.2\% &  0.03 \\
rpoc\_xits\_15\_SAT      & 18   &  97.9\% & 0.03 &  98.8\% & 0.01 & 97.6\% & 0.01 & 99.6\% &  0.16 \\
SAT\_dat.k90.debugged    & 509  &  70.6\% & 8.66 &  68.9\% & 0.53 & 68.8\% & 8.71 & 87.5\% &  18.3 \\
slp-synthesis-aes-top28  & 27   &  64.3\% & 0.53 &  64.2\% & 0.01 & 64.2\% & 0.23 & 98.7\% &  0.10 \\
velev-vliw-uns-2.0-uq5   & 247  &  94.2\% & 0.85 &  93.5\% & 0.35 & 93.2\% & 0.18 & 96.5\% &  4.18 \\
9vliw\_m\_9s\_iq3\_C1\_b1  & 1338 &  86.0\% & 4.13 &  82.3\% & 2.27 & 82.4\% & 1.58 & 86.6\% &  266  \\

\hline
\end{tabular}
\end{center}
\end{table}

\begin{table}
\caption{ We run \emph{PureEager} and \emph{MixDecompose} on 297
application instances. Due to limited space and the fact that
listing all is tedious, we list results on only a part of
application ones and a random instance in the last row. Time is in
seconds.}
\begin{center}

\setlength\tabcolsep{4pt}
\begin{tabular}{|l|r|c|c|c|c|}
\hline  \hline
\  & & \multicolumn{2}{c|} {\emph{PureEager}} &  \multicolumn{2}{c|}{\emph{MixDecompose}} \\
\cline{3-6} \multicolumn{1}{|c|}{\raisebox{1.5ex}[0pt]{Instances}}
  & \raisebox{1.0ex}[0pt]{\large $\frac{|F|}{10^4}$} & $\frac{|L|}{|F|}$ & Time & $\frac{|L|}{|F|}$ & Time \\
 \hline
002-23-96 & 13 & 97.7\% & 1.4 & 99.3\% & 0.29 \\
aes\_24\_4\_keyfind\_4 & 1 & 57.5\% & 0.02 & 68\% & 0.11 \\
atco\_enc1\_opt1\_03\_56 & 26 & 79.3\% & 0.43 & 83.5\% & 7.89\\
blocks-blocks-36-0.120 & 607 & 92.3\% & 17.4 & 96.4\% & 12.65 \\
complete-500-0.1-17 & 8 & 93.9\% & 2.2 & 96.4\% & 3.04 \\
dated-10-11-u  &  49 & 81.6\% & 1.43 & 82.6\% & 2.91 \\
dimacs   & 1 & 99.9\% & 0.14 & 99.9\% & 0.04 \\
grid-strips-grid-y-3.035 & 167 & 85.1\% & 5.61 & 95.1\% & 4.18 \\
hitag2-7-60-0-80 & 3 & 73.8\% & 0.26 & 98.4\% & 1.02\\
MD5-29-3 & 7 & 81.4\% & 0.29 & 99.3\% & 0.51 \\
openstacks-p30\_3.085 & 141 & 93.5\% &1.73 & 94\% & 3.62 \\
partial-5-17-s & 101  & 74.5\% & 2.3 & 82.1\% & 5.66 \\
q\_query\_3\_L150\_coli.sat & 217 & 67.9\% & 52.4 & 85.8\% & 12.03 \\
q\_query\_3\_L90\_coli.sat & 118 & 67.8\% & 15.5 & 88.1\% & 9.04 \\
9vliw\_m\_9stage\_iq3\_C1\_b7 & 1338 &  & $>300$ & 86.8\% & 108.8 \\
9vliw\_m\_9stage\_iq3\_C1\_b4 & 1335 &  & $>300$ & 86.7\% & 109.2 \\

9dlx\_vliw\_at\_b\_iq6 &  364 & 95.2\% & 14.7 & 96.6\% & 12.05\\
SAT\_dat.k75-24\_1\_rule\_3 & 415 & 78.9\% & 14.4 & 87.8\% & 33.83 \\
transport-35node-1000s-4d & 590 & 92.5\% & 24.4 & 92.9\% & 15.66 \\

7pipe\_k &  75 &  0.97.0\% &  1.92  & 99.9\%  &  11.75 \\
ACG-15-10p1 &  94 &  76.4\% & 3.34  & 79.6\% & 7.07 \\

ctl\_3791\_556\_unsat &  8 & 89.0\% & 0.22 & 93.6 & 3.43\\

korf-18  & 19 & 99.3\% & 3.61 & 99.7\% &  10.79 \\
E02F22   & 130 & 99.6\% & 125.8 & 99.9\% &  30.92 \\
MD5-30-4 & 7 & 86.1\% & 0.46  & 99.3\% &  0.83 \\
partial-10-11-s & 68 & 74.6\% & 1.99 & 83.1\%  & 4.78 \\
rbcl\_xits\_08\_UNSAT  &  7 & 99.7\%  & 0.17  & 99.8\% & 0.12 \\
stable-400-0.1-4 & 3 & 91.2\% & 0.49 & 94.4\% & 3.89 \\
total-10-13-u &  79 & 80.9\% &  2.94 & 81.9\% &  7.60 \\
UCG-15-10p0   &  79 & 71.0\% & 3.38  & 78.0\% &  5.76 \\
UR-20-10p1    &  113 & 70.6\% &  4.84 & 76.3\% & 7.85 \\
UTI-20-5p1    &  99  & 70.6\% &  4.24 & 76.5\% & 13.9 \\
velev-vliw-uns-4.0-9-i1 &  323 & 81.3\% &  34.29 & 86.5\% & 21.04\\
IBM\_FV\_2004\_SAT\_dat.k40 & 18 & 91.9\% & 0.83 & 96.4\% & 4.22\\

unif-k3-r3.96-v1000000-c3960000 &     &        &        &   & \\

S8043316035928452744 & \raisebox{1.1ex}[0pt]{396} &
\raisebox{1.1ex}[0pt]{76.3\%} & \raisebox{1.1ex}[0pt]{37.96}
& \raisebox{1.1ex}[0pt]{83.2\%} &\raisebox{1.1ex}[0pt]{82.18}\\

 \hline
\end{tabular}
\end{center}
\end{table}

\begin{table}
\caption{Comparing performance of two algorithms on 297 benchmarks
from SAT Competition 2014 application track.}
\begin{center}

\setlength\tabcolsep{4pt}
\begin{tabular}{|l|c|c|c|c|c|c|}
\hline  \hline
\multicolumn{1}{|c|}{Algorithm} & Ave $\frac{|L|}{|F|}$ & \# of best & \# of eq & Ave Time & Time Out \\
\hline
\emph{PureEager} &  87.2\% & 0 & 71 & 7.41 & 7 \\
\hline
\emph{MixDecompose} &  92.2\% &226 & 71 & 8.97 & 0 \\
 \hline
\end{tabular}
\end{center}
\end{table}

To evaluate the performance of \emph{MixDecompose}, we select very
competitive \emph{PureDecompose}+\emph{EagerMover} (\emph{PureEager}
for short) \cite{web:14,EagerMover:14} as our comparison object.
Although \emph{QuickDecompose} \cite{sbliter:13} was proposed
recently also,  we did not select it as as our comparison object,
because \emph{QuickDecompose} requires more time than
 \emph{EagerMover} for many instances.

   The large set $L$ obtained by \emph{PureEager} contains blockable clauses in addition to
 blocked clauses. A clause $C$ is said to be blockable w.r.t. a blocked set $L$
 if each literal $l \in C$ is not a blocking literal of any clause in $L$.
The reason why blockable clauses are added to the blocked set is
that they do not destroy the blocked property. That is, blocked sets
containing blockable clauses are still satisfiable. To keep
identical with the performance evaluation
  of \emph{PureEager}, the large set $L$ of our
  \emph{MixDecompose} contains also blockable clauses.

Table\,4 compares the performance of \emph{PureEager} and
\emph{MixDecompose} on application instances and a random instance
from the SAT competition 2014. Although we tested the two algorithms
on 297 application instances,  due to limited space and the fact
that listing all yields a tedious feeling, Table\,4 lists only a
part of representative results.  As seen in Table\,4, in terms of
decomposition quality, \emph{MixDecompose} outperforms completely
\emph{PureEager}.  In terms of speed, the former
 is sometimes faster than the latter, and vice versa. \emph{MixDecompose}
 was able to finish the decomposition on all SAT 2014 application
 benchmarks excluding three huge instances within 110 seconds.
 However, \emph{PureEager} was not able to finish on some
 benchmarks such as \emph{9vliw\_m\_9stage\_iq3\_C1\_b7} within 300
 seconds.

Table\,5 presents the outline of the performance of two algorithms
on 297 benchmarks from SAT Competition 2014 application track. The
second column shows the average fraction of the large set. Column
`\# of best' indicates the number of the best results obtained by an
algorithm. Column `\# of eq' is the number of results equivalent to
ones obtained by another algorithm. On 226 out of 297 benchmarks,
the size of the large set obtained by \emph{MixDecompose} is larger
than that obtained by \emph{PureEager}. On 71 remaining benchmarks,
the quality of the two algorithms is identical. There is no
application formula where the quality of \emph{PureEager} is better
than \emph{MixDecompose}. In addition, we conducted also experiments
on random benchmarks. We observed that on all random instances, the
quality of \emph{MixDecompose} is strictly better than that of
\emph{PureEager}. As seen from the last row of Table\,4, \emph{
MixDecompose} can solve huge random instances with millions of
clauses in a reasonable time. For 3-SAT random instances, it can
increase the fraction of the large set by 5\%.

 The fifth column in Table\,5 shows the average runtime taken by each algorithm
 in seconds. Here, computing the average runtime counts only solved instances,
 excluding timed-out instances. The last column in Table\,5, lists the number of times the time-out was hit.
 The timeout for each algorithm was set to 300 seconds. \emph{MixDecompose} did
 not time out on the tested benchmarks, while \emph{PureEager} did
 on 7 benchmarks. \emph{MixDecompose} took at most 110 seconds. Although on average, \emph{PureEager} run faster than
 \emph{MixDecompose}, in this experiment, the worst-case runtime of the former was significantly lager than
 the latter.

\section {Conclusions and Future Work}

  In this paper, we developed a new blocked clause decomposition
algorithm by combining several decomposition strategies. The new
algorithm not only achieves high quality decomposition, but also is
fast. Even for large instances, it can ensure that the decomposition
is done within 110 seconds on our machine. Because our machine is
slower than the platform of SAT competition 2014, If running on the
latter, the speed will be more fast.

 In designing the blocked clause decomposition
algorithm, we simplified Blocked Clause Elimination (BCE) by
applying various cut-off heuristics, such as only "touching" the
literals with few occurrences. We believe that the simple and
limited BCD may be also applied to improve the performance of BCE
for CNF preprocessing, without sacrificing much the quality of the
final result.

   So far we know only that we can get a higher quality decomposition
than the existing algorithms such as \emph{PureEager}. However, this
 does not mean that \emph{MixDecompose} is the best.  How to develop a better
and more efficient than \emph{MixDecompose} will be a future
research topic.

\bibliographystyle{splncs}
\bibliography{fastBCD}

\end{document}